\documentclass[prl,article,preprint,groupedaddress,12pt]{revtex4}
\usepackage{epsfig}
\usepackage{graphicx}
\usepackage{amssymb}

\begin{document}
\title{Black holes that are too cold to respect cosmic censorship}
\author{Shahar Hod}
\affiliation{The Ruppin Academic Center, Emeq Hefer 40250, Israel}
\affiliation{ } \affiliation{The Hadassah Institute, Jerusalem
91010, Israel}
\date{\today}
\centerline {\it This essay is awarded 4th Prize in the 2023 Essay Competition of the Gravity Research Foundation}

\begin{abstract}
\ \ \ In this essay it is proved that there are black holes that are dangerously cold. 
In particular, by analyzing the emission spectra of highly charged black holes we 
reveal the fact that near-extremal black holes whose Bekenstein-Hawking temperatures 
lie in the regime $T_{\text{BH}}\lesssim m^6_e/e^3$ may turn into horizonless naked
singularities, thus violating the cosmic censorship principle, if they emit a photon with the 
characteristic thermal energy $\omega=O(T_{\text{BH}})$ [here $\{m_e,e\}$ are respectively the proper 
mass and the electric charge of the electron, the lightest charged particle]. 
We therefore raise here the conjecture that, in the yet unknown 
quantum theory of gravity, the temperatures of well behaved black-hole spacetimes are fundamentally 
bounded from below by the relation $T_{\text{BH}}\gtrsim m^6_e/e^3$.
\newline
\newline
Email: shaharhod@gmail.com
\end{abstract}
\bigskip
\maketitle

\subsection{Introduction}

The mathematically elegant singularity theorems of Hawking and Penrose
\cite{HawPen,Pen2} have questioned the utility of Einstein's theory of general relativity in 
describing gravitational phenomena in highly curved spacetimes. 
In particular, the Einstein field equations are known to lose their predictive power in the presence of 
infinitely curved regions that contain spacetime singularities.

Following this intriguing observation and in order to guarantee the deterministic nature of 
a self-consistent theory of gravity, Penrose conjectured that a mysterious (and diligent)
``cosmic censor'' protects far away observers from being exposed to 
the pathological properties of spacetime singularities \cite{Pen2}. 
This physically important principle asserts, in particular, that spacetime singularities are always 
hidden inside of black holes with stable shielding horizons. 
If true, the cosmic censorship principle would guarantee 
that we live in a spacetime region in which general relativity is a self-consistent 
theory of gravity \cite{Pen2}.

The most studied curved spacetime in the physics literature, the Kerr-Newman spacetime, describes 
a black hole of mass $M$, electric charge $Q$, and angular momentum $J=Ma$ 
that contains an hidden singularity. 
The characteristic inequality (we use natural Planck units 
in which $G=c=\hbar=k_{\text{B}}=4\pi\epsilon_0=1$) \cite{Chan,NoteQQ}
\begin{equation}\label{Eq1}
M^2-Q^2-a^2\geq0\
\end{equation}
provides a necessary condition for the existence of an engulfing event horizon that protects far away observers 
from being exposed to this inner spacetime singularity. Extremal black-hole configurations, which satisfy the critical 
relation $M^2-Q^2-a^2=0$, are on the verge of exposing their inner singularities. 

In the present essay we shall explicitly prove that well behaved charged black holes 
that respect the condition (\ref{Eq1}) may turn into horizonless naked singularities that violate 
the cosmic censorship principle if they quantum mechanically emit 
massless photons whose characteristic energies are of the same order of magnitude as the thermal energy 
of the black hole.

\subsection{Hawking evaporation of near-extremal charged black holes}

We shall now analyze the physical and mathematical properties of the 
Hawking emission spectra of near-extremal charged black holes whose Hawking-Bekenstein 
temperatures are characterized by the strong dimensionless inequality \cite{Bek1,Haw1}
\begin{equation}\label{Eq2}
\epsilon\equiv 2\pi MT_{\text{BH}}={{M(M^2-Q^2)^{1/2}}\over{[M+(M^2-Q^2)^{1/2}]^2}}\ll1\  .
\end{equation}

The emission rate of neutral bosonic fields from the black hole is given by the familiar 
Hawking relation \cite{Haw1}
\begin{equation}\label{Eq3}
{{dN}\over{dt}}={{1}\over{2\pi}}\sum_{l,m}\int_0^{\infty} {{S_{lm}(\omega)}\over{e^{\omega/T_{\text{BH}}}-1}}d\omega\  ,
\end{equation}
where $\{l,m\}$ are the spheroidal and azimuthal angular harmonic indices of the radiated field mode. 
The partial back-scattering of the emitted fields by the effective 
curvature-centrifugal barrier outside the black hole is encoded in the energy-dependent absorption probability 
factor $S_{lm}(\omega)$ \cite{Haw1}. 

Interestingly, and most importantly for our analysis, it has been explicitly proved in \cite{Pagec} that 
the $l$-dependent centrifugal barrier outside the black hole mainly affects (blocks) the propagation of high-$l$ modes. 
Thus, the Hawking spectra of spherically symmetric large-mass 
black holes \cite{Notelrg} are dominated by the emission of massless field modes with the 
smallest known angular momentum: namely, by electromagnetic field quanta with $l=1$ \cite{Notel1}.

The appearance of both a thermal factor
$(e^{\omega/T_{\text{BH}}}-1)^{-1}$ and an absorption factor $S_{lm}(\omega)$ 
in the black-hole emission formula (\ref{Eq3}) implies that 
the radiation spectrum has a well defined peak which is characterized by the relation 
(below we shall determine the exact value of the dimensionless ratio $\omega_{\text{peak}}/T_{\text{BH}}$)
\begin{equation}\label{Eq4}
\omega_{\text{peak}}\sim T_{\text{BH}}\ll1\  .
\end{equation}
The energy-dependent absorption probability factor of the dominant electromagnetic field mode with $l=1$ is known 
in a closed analytical form in the characteristic regime (\ref{Eq4}) \cite{Pagec,Hodec}: 
\begin{equation}\label{Eq5}
S_{1m}(\nu)={256\over9}\epsilon^8(4\nu^8+5\nu^6+\nu^4)\cdot[1+O(\nu\epsilon)]\  ,
\end{equation}
where we have used here the dimensionless frequency-temperature parameter
\begin{equation}\label{Eq6}
\nu\equiv {{\omega}\over{4\pi T_{\text{BH}}}}\  .
\end{equation}

Substituting the relation (\ref{Eq5}) into the integral expression (\ref{Eq3}) for the 
black-hole radiation rate, one obtains the remarkably compact near-extremal relation \cite{Notemd}
\begin{equation}\label{Eq7}
{{dN_{\gamma}}\over{dt}}={{512\epsilon^{9}}\over{3\pi M}}\int_0^{\infty} {\cal N}(\nu)d\nu\  ,
\end{equation}
where
\begin{equation}\label{Eq8}
{\cal N}(\nu)\equiv{{(1+\nu^2)(1+4\nu^2)\nu^4}\over{e^{4\pi\nu}-1}}\  .
\end{equation}

From Eq. (\ref{Eq8}) one deduces that the emission spectra of black holes 
in the dimensionless near-extremal regime (\ref{Eq2}) are characterized by 
a well defined peak at $\nu_{\text{peak}}\simeq 0.399$. 
Thus, the characteristic energy of an emitted field quantum is given by the functional relation [see Eq. (\ref{Eq6})]
\begin{equation}\label{Eq9}
E=\omega=\nu_{\text{peak}}\cdot4\pi T_{\text{BH}}\  .
\end{equation}

The emission of neutral field modes from near-extremal black holes is dangerous 
from the point of view of the cosmic censorship principle since it allows the emitting 
black hole to reduce its mass without reducing its electric charge. 
The emission of neutral field quanta therefore decreases the magnitude 
of the expression $M^2-Q^2-a^2$ that appears in the necessary condition (\ref{Eq1}) for the existence 
of a shielding horizon that protects far away observers from being exposed to 
the pathological properties of the inner black-hole singularity. 

In particular, the emission of a photon with the characteristic energy (\ref{Eq9}) 
and an azimuthal angular momentum $m$ (with $|m|\in\{0,1\}$) produces a new spacetime configuration 
whose physical parameters are characterized by the following relations:
\begin{equation}\label{Eq10}
M_{\text{new}}=M-E\ \ \ \ ;\ \ \ \ Q_{\text{new}}=Q\ \ \ \ ;\ \ \ \ |a|_{\text{new}}={{|m|}\over{M-E}}\  .
\end{equation}

Intriguingly, taking cognizance of the necessary condition (\ref{Eq1}) for the existence of a black-hole horizon that 
covers the central singularity, one deduces from (\ref{Eq9}) and (\ref{Eq10}) that a near-extremal black hole 
in the dimensionless low-temperature regime
\begin{equation}\label{Eq11}
T_{\text{BH}}<T^{\text{critical}}_{\text{BH}}\equiv{{{\cal C}}\over{\pi M^3}}\
\end{equation}
with ${\cal C}\equiv\nu_{\text{peak}}+\sqrt{\nu^2_{\text{peak}}+{1/4}}$ 
that emits a photon with the characteristic energy (\ref{Eq9}) and angular momentum $l=|m|=1$ leaves behind it 
an horizonless naked singularity that violates the black-hole condition (\ref{Eq1}). 

\subsection{Huge, cold black holes endanger the cosmic censorship principle}

The intriguing conclusion that black holes in the regime (\ref{Eq11}) are too cold to respect cosmic censorship 
is based on our assumption that the radiation spectra of 
near-extremal black holes are dominated by the emission of neutral massless field modes (mainly by photons with $l=1$). 
In particular, since the physical parameters of the positron, the lightest charged particle of the Standard Model, 
are characterized by the strong inequality $e\gg m_e$, even a single positron emission 
would push a near-extremal black hole away from the 
dangerous extremal limit by increasing its temperature (\ref{Eq2}).  

As we shall now prove explicitly, there exists a critical black-hole mass, $M=M_{\text{min}}$, above which 
the radiation spectra of near-extremal black holes are dominated by the emission of neutral massless field modes 
with the smallest known angular momentum (that is, by photons with $l=1$ \cite{Pagec}). 
Black holes in the near-extremal regime (\ref{Eq11}) with $M>M_{\text{min}}$ may evaporate into horizonless 
naked singularities that violate the black-hole condition (\ref{Eq1}), thus violating the cosmic censorship principle.  

In order to determine the value of the critical black-hole mass $M_{\text{min}}$, 
one may use the relation [see Eqs. (\ref{Eq2}), (\ref{Eq7}), 
and (\ref{Eq8})]
\begin{equation}\label{Eq12}
{{dN_{\gamma}}\over{dt}}={{256\xi}\over{\pi}}\cdot M^8T^9_{\text{BH}}\
\end{equation}
with $\xi\equiv 8\pi^4\zeta(5)+75\pi^2\zeta(7)+210\zeta(9)\simeq 1764.9$ 
for the emission rate of neutral massless photons with $l=1$ from near-extremal black holes in the 
dimensionless regime (\ref{Eq11}). 
In addition, one may use the fact that, 
in the regime $1\ll Mm_e\ll Qe\ll (Mm_e)^2$, 
the emission rate of charged field quanta (positrons) by near-extremal black holes is well approximated by the Schwinger 
pair-production formula \cite{Pagec,Sch1,Sch2}
\begin{equation}\label{Eq13}
{{dN_{e^+}}\over{dt}}={{e^3}\over{2\pi^3m^2_e}}\cdot \exp({-E_{\text{c}}/E_+})\  ,
\end{equation}
where $E_+=Q/r^2_+\simeq 1/Q$ is the electric field strength of the near-extremal black hole and 
$E_{\text{c}}=\pi m^2_e/e$ is the critical electric field for quantum production of electron-positron pairs.

Our assumption that the emission spectrum of 
the near-extremal black hole with the critical temperature (\ref{Eq11}) 
is dominated by neutral massless photons with $l=1$ corresponds to the relation
\begin{equation}\label{Eq14}
{{dN_{e^+}}\over{dt}}<{{dN_{\gamma}}\over{dt}}\  .
\end{equation}
Taking cognizance of Eqs. (\ref{Eq11}), (\ref{Eq12}), and (\ref{Eq13}), one can express  
the inequality (\ref{Eq14}) in the form
\begin{equation}\label{Eq15}
\Big({{\pi m^2_e}\over{e}}M\Big)^{19}\cdot\exp\Big(-{{\pi m^2_e}\over{e}}M\Big)<
{{512\xi{\cal C}^9}\over{\pi^8e^2}}\cdot\Big({{\pi m^2_e}\over{e}}\Big)^{20}\  ,
\end{equation}
which yields the critical relation (note that, in natural Planck units, the physical parameters of the positron 
are given by $e\simeq 1/137.036^{1/2}$ 
and $m_e\simeq 4.19\cdot 10^{-23}$) 
\begin{equation}\label{Eq16}
M>M_{\text{min}}\equiv{{e}\over{\pi m^2_e}}\cdot x_{\text{min}}\simeq 1.55\times10^{43}\cdot x_{\text{min}}\
\end{equation}
with $x_{\text{min}}\simeq 2124.7$ \cite{Notems}. 

The Hawking radiation spectrum of the near-extremal black hole with the 
critical temperature $T^{\text{critical}}_{\text{BH}}={{{\cal C}}/{\pi M^3}}$ [see Eq. (\ref{Eq11})] 
is dominated, in the large-mass regime (\ref{Eq16}), by the emission of neutral massless photons that 
endanger the integrity of the black-hole horizon.

\subsection{Summary and discussion}

The Penrose cosmic censorship principle asserts that general relativity is a
deterministic theory of gravity and that pathological spacetime singularities are always 
hidden inside of black holes with stable shielding horizons \cite{HawPen,Pen2}. 

In the present essay we have explicitly proved that near-extremal (cold and large) black holes 
may evaporate into naked singularities that violate the cosmic censorship principle.
In particular, taking cognizance of the analytically derived relations (\ref{Eq11}) and (\ref{Eq16}), 
one deduces that the threat to the validity of the principle is limited to the extreme physical regime
\begin{equation}\label{Eq17}
T_{\text{BH}}<T^{\text{critical}}_{\text{BH}}={{{\cal C}}\over{\pi x^3_{\text{min}}}}\cdot
\Big({{\pi m^2_e}\over{e}}\Big)^3\simeq 10^{-140}\simeq 10^{-108}{\ ^\circ K}\
\end{equation}
of ultra-cold black holes. 

Since we believe that cosmic censorship should be one of the cornerstones of a self-consistent theory of gravity 
in curved spacetimes, we here raise the conjecture that, in the yet unknown 
quantum theory of gravity, the temperatures of well behaved black-hole spacetimes are fundamentally 
bounded from below by a relation of the form
\begin{equation}\label{Eq18}
T_{\text{BH}}\gtrsim{{m^6_e}\over{e^3}}\  .
\end{equation}
If the lower bound (\ref{Eq18}) is indeed respected, 
then the emission of characteristic quanta from near-extremal black holes would not 
endanger the validity of cosmic censorship, a principle which is fundamentally important for a self-consistent 
formulation of the microscopic quantum theory of gravity.  

\newpage

\bigskip
\noindent
{\bf ACKNOWLEDGMENTS}
\bigskip

This research is supported by the Carmel Science Foundation. I thank
Don Page for interesting correspondence. I would also like to thank
Yael Oren, Arbel M. Ongo, Ayelet B. Lata, and Alona B. Tea for
stimulating discussions.



\begin{thebibliography}{99}
\bibitem{NoteQQ} We shall henceforth assume, without loss of generality, the relation $Q>0$ 
for the electric charge of the black hole.
\bibitem{Notelrg} In Eq. (\ref{Eq16}) below we shall provide a precise definition for the concept 
of ``large-mass black holes'' in the context of black-hole evaporation.
\bibitem{Notel1} As nicely discussed in \cite{Pagec}, the presence of the $l$-dependent centrifugal
barrier around the radiating black hole implies that, as compared to the emission rate of photons 
with the minimally allowed value $l=1$ of the spheroidal angular index, 
the emission rates of massless fields 
with $l>1$ to infinity are suppressed by several factors of the large dimensionless quantity $1/(MT_{\text{BH}})$.
\bibitem{Notemd} Here we have taken into account the three possible modes $m=-1,0,1$ for the 
electromagnetic field with $l=1$ and the two possible polarizations for each mode.
\bibitem{Notems} Note that the critical black-hole mass (\ref{Eq16}) corresponds 
to $M_{\text{min}}\simeq 3.6\times 10^8M_{\odot}$.
\end{thebibliography}
\end{document}